\documentclass[prl, superscript address,float fix,one column,reprint,footinbib]{revtex4-1}

\usepackage{amsmath,amssymb,bm,epsfig}
\usepackage{color}
\usepackage{natbib}
\usepackage{hyperref} 
\usepackage{ulem}
\usepackage{graphicx}
\usepackage{wasysym}
\usepackage{cleveref}
\RequirePackage{lineno}

\usepackage[utf8]{inputenc}

\usepackage{xcolor,colortbl}
\usepackage{pifont}
\usepackage{footnote}

\def \beq{\begin{equation}}
\def \eeq{\end{equation}}
\def \beqa{\begin{eqnarray}}
\def \eeqa{\end{eqnarray}}
\def \la{\langle}
\def \ra{\rangle}

\begin{document}
\date{\today}
\title{Constraining the initial conditions and temperature dependent transport with three-particle correlations in Au+Au collisions}


\affiliation{AGH University of Science and Technology, FPACS, Cracow 30-059, Poland}
\affiliation{Argonne National Laboratory, Argonne, Illinois 60439}
\affiliation{Brookhaven National Laboratory, Upton, New York 11973}
\affiliation{University of California, Berkeley, California 94720}
\affiliation{University of California, Davis, California 95616}
\affiliation{University of California, Los Angeles, California 90095}
\affiliation{Central China Normal University, Wuhan, Hubei 430079}
\affiliation{University of Illinois at Chicago, Chicago, Illinois 60607}
\affiliation{Creighton University, Omaha, Nebraska 68178}
\affiliation{Czech Technical University in Prague, FNSPE, Prague, 115 19, Czech Republic}
\affiliation{Nuclear Physics Institute AS CR, 250 68 Prague, Czech Republic}
\affiliation{Frankfurt Institute for Advanced Studies FIAS, Frankfurt 60438, Germany}
\affiliation{Institute of Physics, Bhubaneswar 751005, India}
\affiliation{Indiana University, Bloomington, Indiana 47408}
\affiliation{Alikhanov Institute for Theoretical and Experimental Physics, Moscow 117218, Russia}
\affiliation{University of Jammu, Jammu 180001, India}
\affiliation{Joint Institute for Nuclear Research, Dubna, 141 980, Russia}
\affiliation{Kent State University, Kent, Ohio 44242}
\affiliation{University of Kentucky, Lexington, Kentucky, 40506-0055}
\affiliation{Lamar University, Physics Department, Beaumont, Texas 77710}
\affiliation{Institute of Modern Physics, Chinese Academy of Sciences, Lanzhou, Gansu 730000}
\affiliation{Lawrence Berkeley National Laboratory, Berkeley, California 94720}
\affiliation{Lehigh University, Bethlehem, PA, 18015}
\affiliation{Max-Planck-Institut fur Physik, Munich 80805, Germany}
\affiliation{Michigan State University, East Lansing, Michigan 48824}
\affiliation{National Research Nuclear University MEPhI, Moscow 115409, Russia}
\affiliation{National Institute of Science Education and Research, Bhubaneswar 751005, India}
\affiliation{National Cheng Kung University, Tainan 70101 }
\affiliation{Ohio State University, Columbus, Ohio 43210}
\affiliation{Institute of Nuclear Physics PAN, Cracow 31-342, Poland}
\affiliation{Panjab University, Chandigarh 160014, India}
\affiliation{Pennsylvania State University, University Park, Pennsylvania 16802}
\affiliation{Institute of High Energy Physics, Protvino 142281, Russia}
\affiliation{Purdue University, West Lafayette, Indiana 47907}
\affiliation{Pusan National University, Pusan 46241, Korea}
\affiliation{Rice University, Houston, Texas 77251}
\affiliation{University of Science and Technology of China, Hefei, Anhui 230026}
\affiliation{Shandong University, Jinan, Shandong 250100}
\affiliation{Shanghai Institute of Applied Physics, Chinese Academy of Sciences, Shanghai 201800}
\affiliation{State University Of New York, Stony Brook, NY 11794}
\affiliation{Temple University, Philadelphia, Pennsylvania 19122}
\affiliation{Texas A\&M University, College Station, Texas 77843}
\affiliation{University of Texas, Austin, Texas 78712}
\affiliation{University of Houston, Houston, Texas 77204}
\affiliation{Tsinghua University, Beijing 100084}
\affiliation{University of Tsukuba, Tsukuba, Ibaraki, Japan,}
\affiliation{Southern Connecticut State University, New Haven, CT, 06515}
\affiliation{United States Naval Academy, Annapolis, Maryland, 21402}
\affiliation{Valparaiso University, Valparaiso, Indiana 46383}
\affiliation{Variable Energy Cyclotron Centre, Kolkata 700064, India}
\affiliation{Warsaw University of Technology, Warsaw 00-661, Poland}
\affiliation{Wayne State University, Detroit, Michigan 48201}
\affiliation{World Laboratory for Cosmology and Particle Physics (WLCAPP), Cairo 11571, Egypt}
\affiliation{Yale University, New Haven, Connecticut 06520}

\author{L.~Adamczyk}\affiliation{AGH University of Science and Technology, FPACS, Cracow 30-059, Poland}
\author{J.~K.~Adkins}\affiliation{University of Kentucky, Lexington, Kentucky, 40506-0055}
\author{G.~Agakishiev}\affiliation{Joint Institute for Nuclear Research, Dubna, 141 980, Russia}
\author{M.~M.~Aggarwal}\affiliation{Panjab University, Chandigarh 160014, India}
\author{Z.~Ahammed}\affiliation{Variable Energy Cyclotron Centre, Kolkata 700064, India}
\author{N.~N.~Ajitanand}\affiliation{State University Of New York, Stony Brook, NY 11794}
\author{I.~Alekseev}\affiliation{Alikhanov Institute for Theoretical and Experimental Physics, Moscow 117218, Russia}\affiliation{National Research Nuclear University MEPhI, Moscow 115409, Russia}
\author{D.~M.~Anderson}\affiliation{Texas A\&M University, College Station, Texas 77843}
\author{R.~Aoyama}\affiliation{University of Tsukuba, Tsukuba, Ibaraki, Japan,}
\author{A.~Aparin}\affiliation{Joint Institute for Nuclear Research, Dubna, 141 980, Russia}
% ...

\collaboration{STAR Collaboration}\noaffiliation

\begin{abstract} 
We present three-particle mixed-harmonic correlations $\la \cos (m\phi_a + n\phi_b - (m+n) \phi_c)\ra$ for harmonics $m,n=1-3$ for charged particles in $\sqrt{s_{NN}}=$200 GeV Au+Au collisions at RHIC. These measurements provide information on the three-dimensional structure of the initial collision zone and are important for constraining models of a subsequent low-viscosity quark-gluon plasma expansion phase. We investigate correlations between the first, second and third harmonics predicted as a consequence of fluctuations in the initial state. The dependence of the correlations on the pseudorapidity separation between particles show hints of a breaking of longitudinal invariance. We compare our results to a number of state-of-the art hydrodynamic calculations with different initial states and temperature dependent viscosities. These measurements provide important steps towards constraining the temperature dependent transport and the longitudinal structure of the initial state at RHIC. 

\end{abstract}


\maketitle

{\it Introduction :}  Matter as hot and dense as the early universe microseconds after the Big Bang can be created by colliding heavy  nuclei at high energies. At these temperatures, baryons and mesons melt to form a quark gluon plasma (QGP)~\cite{PhysRevLett.34.1353, Chin1978552, Kapusta1979461,  PhysRevD.22.2793}. Data from the Relativistic Heavy Ion Collider (RHIC) at Brookhaven National Laboratory and the Large Hadron Collider (LHC) at CERN have been arguably used to show that the QGP at these temperatures is a nearly perfect fluid with a shear viscosity-to-entropy density ratio ($\eta$/s) smaller than any other fluid known in nature~\cite{Arsene:2004fa, Back:2004je, Adams:2005dq, Adcox:2004mh, Muller:2007rs, Zajc:2007ey, Gale:2012rq, Chatrchyan:2013kba, Abelev:2014pua}. Theoretical calculations suggest that like many other fluids, the QGP viscosity should have a dependence on temperature with a minimum at the QGP-to-hadron transition temperature~\cite{Prakash:1993bt, Arnold:2003zc, Csernai:2006zz}. The determination of the temperature dependence of these transport properties is an open problem of fundamental importance in the study of the emerging properties of QCD matter.

Over the past years the harmonic decomposition of two-particle azimuthal correlations $v_n^2\{2\}=\langle\cos n(\phi_a-\phi_b)\rangle$ (where $\phi_{a,b}$ are azimuthal angles of particle momenta)~\cite{Adams:2004bi, Adare:2011tg, Aad:2014vba, Abelev:2014mda, Chatrchyan:2013kba} have already helped shed light on these topics. Hydrodynamic models with different initial conditions and transport parameters have been compared to measurements at RHIC and LHC to constrain the fluid-like property of the medium~\cite{Gale:2013da}. Given their large number of parameters, measurements of multiple observables over a wide energy range have been found to be essential for constraining such models~\cite{Novak:2013bqa, Pratt:2015zsa, Bernhard:2015hxa}. So far however, the temperature dependence of transport parameters like the bulk and shear viscosity are not well constrained by the existing data. %

In this letter, we report on the measurement of three-particle correlations that provide unique ways to constrain the fluid-like properties of the QGP. These new measurements at RHIC extend beyond the conventional two-particle correlations; they help elucidate the three dimensional structure of the initial state, probe the nonlinear hydrodynamic response of the medium, and will help constrain the temperature dependence of the transport parameters. 

We measure three-particle azimuthal correlations using the observables~\cite{Bhalerao:2013ina}
\begin{equation}
\label{eq_cmn}
C_{m,n,m+n}= \la \la \cos(m\phi_a + n \phi_b - (m+n) \phi_c)\ra\ra 
\end{equation}
where the inner average is taken over all sets of unique triplets, and the outer average is taken over all events weighted by the number of triplets in each event. %
The subscripts ``$m,n$" in $C_{m,n,m+n}$ refer to the harmonic number while the subscripts ``$a,b,c$" in $\phi$ refer to the indices of the particles. 
We report on the centrality dependence of $C_{m,n,m+n}$ with combinations of harmonics $(m,n) = (1,1), (1,2), (2,2), (2,3), (2,4)$ and $(3,3)$ for inclusive charged particles in Au+Au collisions at $\sqrt{s_{NN}}=$ 200 GeV. In a longer companion paper~\cite{STAR:3pclong16} we present our measurements at lower energies ($\sqrt{s_{NN}}=$ 62.4-7.7 GeV). 
The $C_{m,n,m+n}$ are related to event-plane correlations like those measured in Pb+Pb collisions at 2.76 TeV~\cite{Aad:2014fla,Jia:2014jca, ALICE:2016kpq}. 
If $v_n$ and $\Psi_n$ denote~\footnote{
\begin{equation} 
\label{eq:2}
  v_n\,e^{i n \Psi_n} =\frac
  {\int p_T\,dp_T\,d\phi\, e^{i n\phi}\,\frac{dN_\mathrm{ch}}{d\eta\,p_Tdp_T\,d\phi}}
  {\int p_Tdp_T\,d\phi\,\frac{dN_\mathrm{ch}}{d\eta\,p_Tdp_T\,d\phi}}.
\end{equation}
where $\frac{dN_\mathrm{ch}}{d\eta\,p_Tdp_T\,d\phi}$ is the single particle distribution. 
} 
anisotropic flow coefficients and their associated event planes~\cite{Voloshin:2008dg}, for $m,n >1$, $C_{m,n,m+n}$ can be approximated as ${\la v_m v_n v_{m+n} \cos(m \Psi_m + n \Psi_n - (m+n) \Psi_{m+n}) \ra}$. Such flow based interpretation is not likely to be applicable in case of $m,n =1$ for which a strong charge dependence has been observed~\cite{Abelev:2009ac, Abelev:2009ad, Adamczyk:2013hsi} and the effects of global momentum conservation may be important~\cite{Borghini:2000cm, Jia:2012hx}.

\begin{figure}[t]
\includegraphics[width=0.5\textwidth]{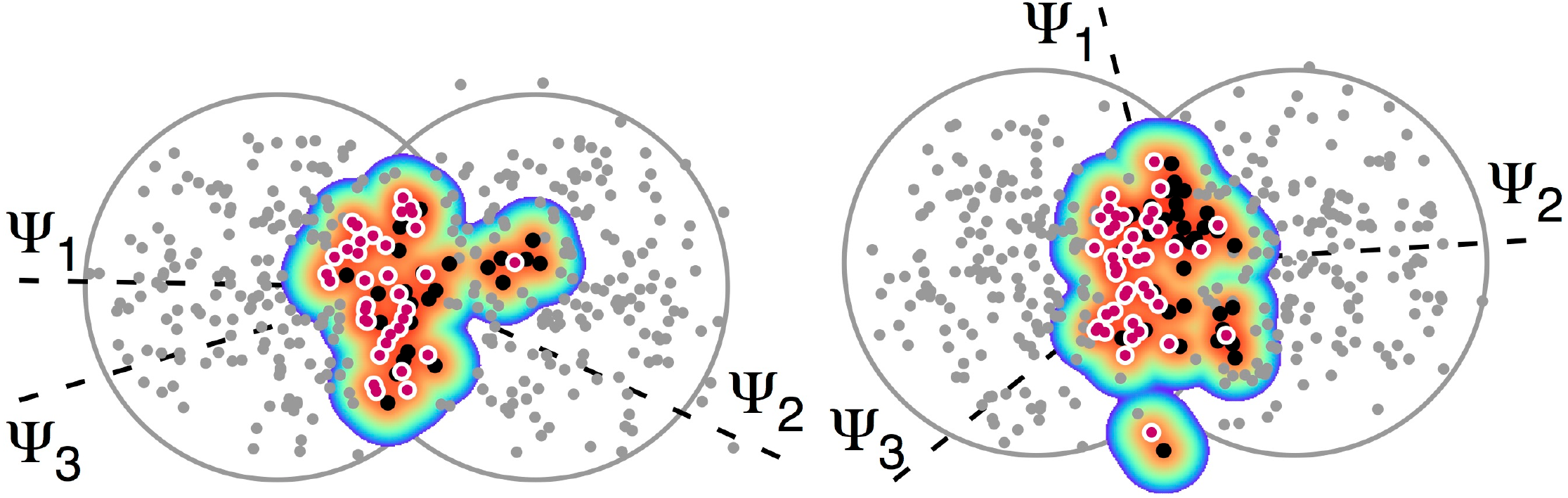}
\caption{(color online) Monte Carlo Glauber simulation for Au+Au collisions at $\sqrt{s_{NN}}$ = 200 GeV with impact parameter $b=10$ fm, showing in-plane and out-of-plane fluctuations of the participants. The grey points show the positions of the spectator nucleons. The positions of the wounded nucleons from the left (right) nucleus are shown by red (black) colored points in each diagram. The Gaussian energy deposition (width = 0.4 fm) around the center of wounded nucleons are shown by color contours. The orientations of different geometric eccentricities are shown by dashed lines.} 
\label{fig_fluc}
\vspace{-10pt}
\end{figure}

Measurements of $C_{m,n,m+n}$ provide unique information about the geometry of the collision overlap region and its fluctuations. Reference ~\cite{Teaney:2010vd} proposed that measurements of $C_{1,2,3}$ could detect event-by-event correlations of the first, second and third harmonic anisotropies. Although it is sometimes assumed that the axis of the third harmonic is random, Monte-Carlo Glauber simulations show correlations between the first, second, and third harmonic planes. Figure \ref{fig_fluc} (left) shows the case when a single nucleon (shown by a red dot) at the edge of a colliding nucleus fluctuates outward and impinges on the other nucleus creating a region of increased energy density. This specific in-plane fluctuation generates $v_1$, which reduces $v_2$ and increases $v_3$~\cite{Adamczyk:2016exq}. 
A similar fluctuation occurring in the out-of-plane direction is illustrated in the right panel of Fig.~\ref{fig_fluc}. 
 Such correlations, if observed in terms of $C_{1,2,3}$, will for the first time, demonstrate the presence of a $v_1$ driven component of $v_3$ arising due to initial geometry. 

The fluctuation illustrated in Fig. \ref{fig_fluc} (left) 
when the nucleon at the edge of one nucleus impinges on the center of the other nucleus, it is similar to a central p+Au collision. In p+Au collisions, the maximum of the multiplicity distribution shifts in pseudorapidity $\eta$ towards the Au going direction. 
For this reason, one expects that the harmonic planes can point in different directions for positive or negative $\eta$. 
Similar effects have been investigated in models and discussed in terms of torqued fireballs~\cite{Bozek:2010vz}, twists~\cite{Jia:2014ysa}, or reaction-plane decorrelations~\cite{Pang:2015zrq}. Studying the $\Delta\eta$ dependence of $C_{1,2,3}$ should reveal these effects if they exist, and provide new insights on the three dimensional structure of the initial state. 

In general, if a medium is fully describable by hydrodynamics, nonlinear couplings between harmonics are expected to change the sign of $C_{m,n,m+n}$ relative to what would be expected based on the initial state eccentricities $\varepsilon_{n}$~\footnote{
\begin{equation}
\varepsilon_n \,e^{i n\Phi_n}
  = -\frac{\int r\,dr\,d\phi\, r^n e^{i n\phi}\,E(r,\phi)}{\int r\,dr\,d\phi\, r^n\,E(r,\phi)}
\end{equation} 
where $E(r,\phi)$ is the distribution of initial energy density.} and participant planes $\Phi_{n}$ ~\cite{Teaney:2010vd, Qiu:2012uy, Teaney:2012ke, Teaney:2013dta, Bhalerao:2013ina, Yan:2015jma, Qian:2016fpi, Qian:2016pau, McDonald:2016vlt, Betz:2016ayq}. Observables sensitive to nonlinear hydrodynamic response are ideal probes of viscosity. Since higher harmonics are more strongly dampened by viscosity, the nonlinear coupling increases correlations of $v_n$ with other lower harmonic eccentricities $\varepsilon_{m<n}$, and thereby with $v_{m< n}$. In this way, $C_{m,n,m+n}$ becomes more sensitive to $\eta/s$ as 
%
%
%
previously demonstrated by phenomenological studies at LHC energies~\cite{Qiu:2012uy,Teaney:2013dta, Bhalerao:2013ina, Niemi:2015qia}. Correlations of event planes and flow harmonics measured by the ATLAS and ALICE collaborations for $m,n\!\ge\!2$~\cite{Aad:2014vba,Jia:2014jca, ALICE:2016kpq} have been compared to hydrodynamic simulations to constrain the temperature dependence of viscosity $\eta/s~(T)$~\cite{Niemi:2015qia}.  However since LHC measurements are sensitive to the $\eta/s$ at higher temperatures, full constraint on $\eta/s~(T)$ is better achieved with 
measurements of observables like $C_{m,n,m+n}$ at RHIC~\cite{Niemi:2011ix, Gale:2012rq, Niemi:2015qia, Denicol:2015nhu}. 

In this work we report the three-particle correlations directly instead of event-plane correlations. Expressing three-particle correlations as event plane correlations relies on factorization, {\it i.e.}, approximations like $C_{m,n,m+n}=\la v_m v_n v_{m+n} \cos(m \Psi_m + n \Psi_n - (m+n) \Psi_{m+n}) \ra =$ $\la v_m \ra \la v_n \ra \la v_{m+n} \ra  \la \cos(m \Psi_m + n \Psi_n - (m+n) \Psi_{m+n})\ra$, that can complicate data-model comparison. We therefore, directly compare $C_{m,n,m+n}$ to theoretical predictions. Another advantage of three-particle correlations is that the measurements are well defined even without assuming the flow coefficients and harmonic planes dominate the correlation. Other effects besides reaction plane correlations, particularly important for $m,n =1$, can be present in $C_{m,n,m+n}$ and the correctness and completeness of a model needs to be judged through direct comparison to the data. 
Also, when the correlations are dominated by reaction plane correlations, $C_{m,n,m+n}$ corresponds to a well-defined limit (the low-resolution limit) ~\cite{Luzum:2012da} of the measurement, which again, makes for a more direct comparison to theory. A more practical advantage is as follows: unlike LHC, since $v_n^2\{2\}$ for $n=1-6$ is not always a large positive quantity at RHIC, it is not always feasible to divide $C_{m,n,m+n}$ by $\sqrt{v_n^2\{2\}}$ to express it purely as an event plane correlation without losing experimental significance. The magnitude of $v_6^2\{2\}$ is negligible at RHIC, $v_5^2\{2\}$ measurements suffer from large systematics, and  $v_1^2\{2\}<0$ except for central events at $\sqrt{s_{NN}}=200$ GeV~\cite{STAR:3pclong16}.

{\it Experiment and Analysis : } We present measurements of $C_{m,n,m+n}$ in 200 GeV Au+Au collisions with data collected in the year 2011 by the STAR detector~\cite{Ackermann2003624} at RHIC. We detect charged particles within the range $|\eta|\!<\!1$ and for transverse momentum of $p_T\!>\!0.2$ GeV/$c$ using the STAR Time Projection Chamber~\cite{Anderson2003659} situated inside a 0.5 Tesla solenoidal magnetic field. We use track-by-track weights~\cite{Bilandzic:2010jr, Bilandzic:2013kga} to account for imperfections in the detector acceptance and momentum dependence of the detector efficiency. We correct the two-track acceptance artifacts which arise due to track-merging effects by measuring the $|\Delta\eta_{ab}|=|\eta_a-\eta_b|$,  $|\Delta\eta_{ac}|=|\eta_a-\eta_c|$, and $|\Delta\eta_{bc}|=|\eta_b-\eta_c|$ dependence of $C_{m,n,m+n}$ and algebraically correcting the integrated value of $C_{m,n,m+n}$ for the missing pairs apparent at $\Delta\eta \approx 0$. 
Note that, throughout this paper, the subscripts ``$m,n$ with comma" in $C_{m,n,m+n}$ refer to the harmonic number while the subscripts ``$ab$ without comma" for the $|\Delta\eta_{ab}|=|\eta_a -\eta_b|$ refer to the indices of the particles. 
We estimate systematic uncertainties by comparing data from different time periods, from different years with different tracking algorithms, by comparing different efficiency estimates, by varying the z-vertex position of the collision, and by varying track selection criteria. We also include estimates of the effect of short-range HBT and Coulomb correlations in the systematic uncertainties based on the shape of the $\Delta\eta$ dependence. For such quantifications we fit the $\Delta\eta$ dependence of $C_{m,n,m+n}$ with the combination of a short-range and a long-range Gaussian distributions as described in Ref~\cite{Tribedy:2017hwn,Adamczyk:2016exq}. Finally, in order to quantify other nonflow effects such correlations due to mini-jets, fragmentation, decay etc. we compare our data to HIJING (Version 1.383) calculations~\cite{Wang:1991hta}. 
For each of our centrality intervals ($0-5\%, 5-10\%, 10-20\%, ..., 70-80\%$), we use a Monte Carlo Glauber model~\cite{Abelev:2008ab,Miller:2007ri} to estimate the average number of participating nucleons $N_{\rm part}$ for  plotting our results~\footnote{See Ref.~\cite{Abelev:2008ab} for details like centrality resolution, values of impact parameter, $N_{\rm part}$ etc.}.

\begin{figure}[t]
\includegraphics[width=0.45\textwidth]{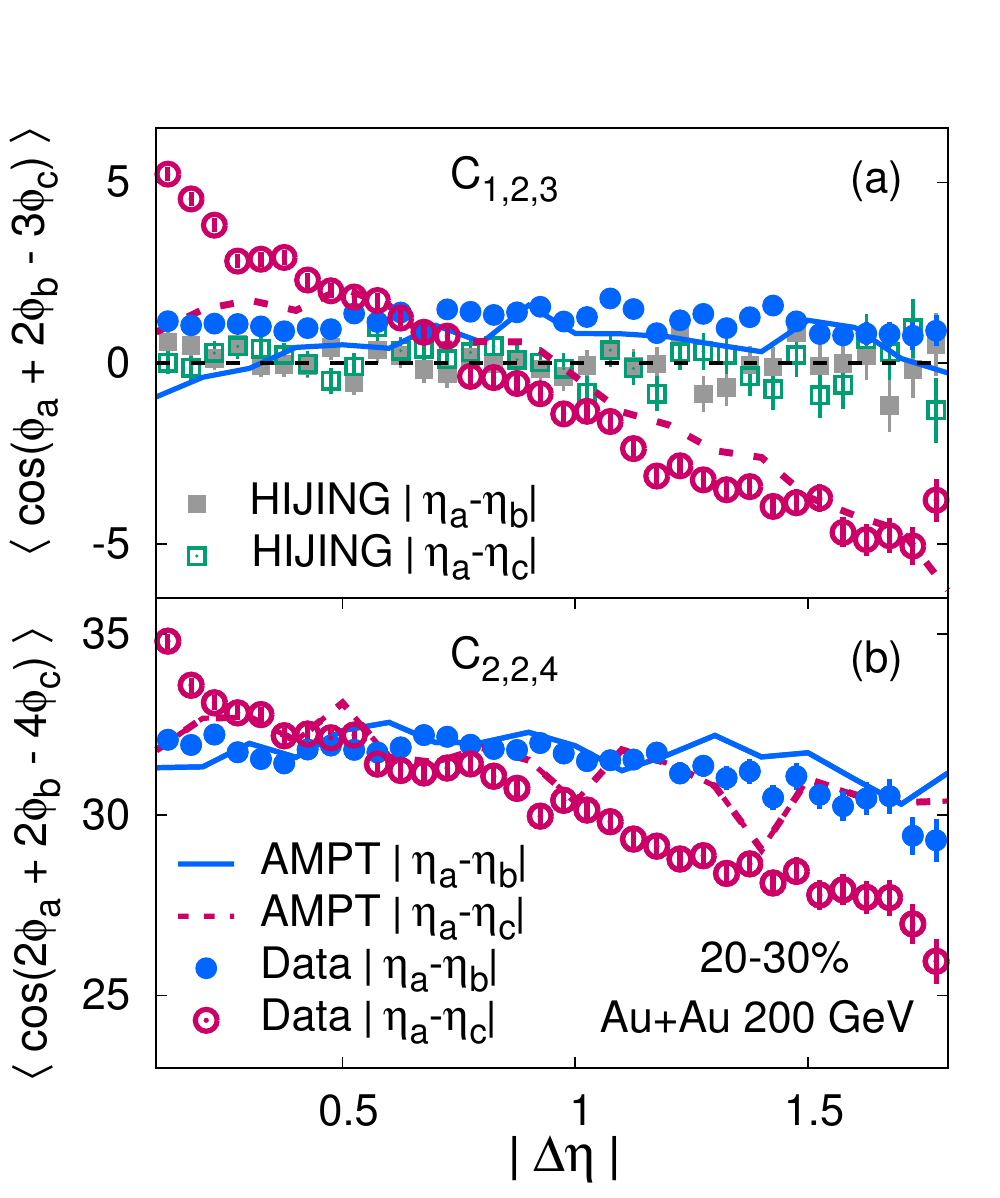}
\caption{(color online) Dependence of mixed harmonic correlators $C_{1,2,3}$ and $C_{2,2,4}$ on relative pseudorapidity. HIJING calculations are shown to quantify short-range nonflow correlations~\cite{Wang:1991hta}. AMPT model~\cite{Lin:2004en} calculations from Ref~\cite{Sun:2017ine} are also compared to demonstrate the effects of 3D initial geometry and transport on three particle correlations.}
\label{fig_cosmn_deta}
\vspace{-10pt} 
\end{figure}

{\it Results : } Figure~\ref{fig_cosmn_deta} (a,b) shows the $\Delta\eta$ dependence of $C_{1,2,3}=\la \cos(\phi_a + 2\phi_b -3\phi_c)\ra$ and $C_{2,2,4}=\la \cos(2\phi_a + 2\phi_b -4\phi_c)\ra$. The $\Delta\eta$ dependence of $C_{1,1,2}=\la \cos(\phi_a + \phi_b -2\phi_c)\ra$ was presented previously~\cite{Abelev:2009ad, Adamczyk:2013hsi} and other harmonic combinations will be presented in Ref.~\cite{STAR:3pclong16}. The top panel of Fig.~\ref{fig_cosmn_deta} shows $C_{1,2,3}$ as a function of $|\Delta\eta_{ab}|$ and $|\Delta\eta_{ac}|$. We observe a strong $|\Delta\eta_{ac}|$ dependence but a weak $|\Delta\eta_{ab}|$ dependence. The observation for $|\Delta\eta_{bc}|$ is similar to $|\Delta\eta_{ab}|$, so we omit it from the figure for clarity. For $|\Delta\eta_{ac}|\approx 0$, $C_{1,2,3}$ is positive, but as $|\Delta\eta_{ac}|$ increases,  $C_{1,2,3}$ decreases and becomes negative. We study the centrality dependence of this effect and find that $C_{1,2,3}$ has the strongest dependence on $|\Delta\eta_{ac}|$ in mid-central events (20-30$\%$); in central (0-5$\%$) and peripheral events (70-80$\%$), $C_{1,2,3}$ shows weaker dependence on $|\Delta\eta_{ac}|$ (see Ref.~\cite{STAR:3pclong16}).  This is consistent with expectations of the breaking of longitudinal invariance through forward-backward rapidity dependence as previously discussed. No such dependence is observed for $|\Delta\eta_{ab}|$ since although the third harmonic plane may rotate significantly in the forward and backward directions, the second harmonic plane should remain invariant due to the symmetry of  collision geometry. 

As mentioned before, since $C_{1,2,3}$ involves the first order harmonic it may have contributions from nonflow correlations such as global momentum conservation~\cite{Borghini:2000cm}. However, such contributions have been argued to be independent of $\Delta\eta$ in leading order~\cite{Borghini:2000cm,Pratt:2010gy,Bzdak:2010fd}. One, therefore, can not explain the strong variation of $C_{1,2,3}$ with $|\Delta\eta_{ac}|$ even up to 2, which is strongest in the mid-central events, to be only as an artifact of momentum conservation.

The HIJING model comparisons shown in Fig.~\ref{fig_cosmn_deta} demonstrate that nonflow contributions due to mini-jets can not explain data. On the other hand the AMPT model~\cite{Lin:2004en} calculations from Ref.~\cite{Sun:2017ine} that involves momentum conservation, mini-jets, as well as collectivity due to multiphase transport, and three-dimensional initial state seem to provide a better description of the $\Delta\eta$ dependence of $C_{1,2,3}$ above $\Delta\eta>0.5$; at smaller $\Delta\eta<0.5$ AMPT under predicts the data.

In Fig.~\ref{fig_cosmn_deta} (b) we present the $\Delta\eta$ dependence of $C_{2,2,4}$. 
We find much weaker $\Delta\eta$ dependence for $C_{2,2,4}$ than for $C_{1,2,3}$; while $C_{1,2,3}$ changes sign, $C_{2,2,4}$ only varies by 20\% over the range of our measurements. This is not surprising since the second harmonic event plane dominates $C_{2,2,4}$. The dependence of $C_{2,2,4}$ is also stronger for $|\Delta\eta_{ac}|$ than it is for $|\Delta\eta_{ab}|$. Once again, the HIJING predictions (not shown in this figure) are much smaller and consistent with zero. The AMPT predictions from Ref~\cite{Sun:2017ine} do a very good job in describing the magnitude of the correlation, it however, seem to slightly under predict the slope of the $\Delta\eta$ dependence. 

We find that all the correlators exhibit a significant $\Delta\eta$ dependence except 
$C_{2,2,4}$ and $C_{2,3,5}$ 
which vary by only $20\%$~\cite{STAR:3pclong16}. The variation of $C_{m,n,m+n}$ with $\Delta\eta$ makes it difficult to compare the data to models that assume a longitudinally invariant two-dimensional (boost invariant) initial geometry. Until those simplifying assumptions are relaxed, $C_{2,2,4}$ and $C_{2,3,5}$ having the smallest relative variation on $\Delta\eta$ provide the best opportunity for comparison of $\Delta\eta$-integrated quantities with hydrodynamic models. 

\begin{figure*}[t]
\includegraphics[trim={0.5cm 0.5cm 0.5cm 0.5cm}, width=1\textwidth]{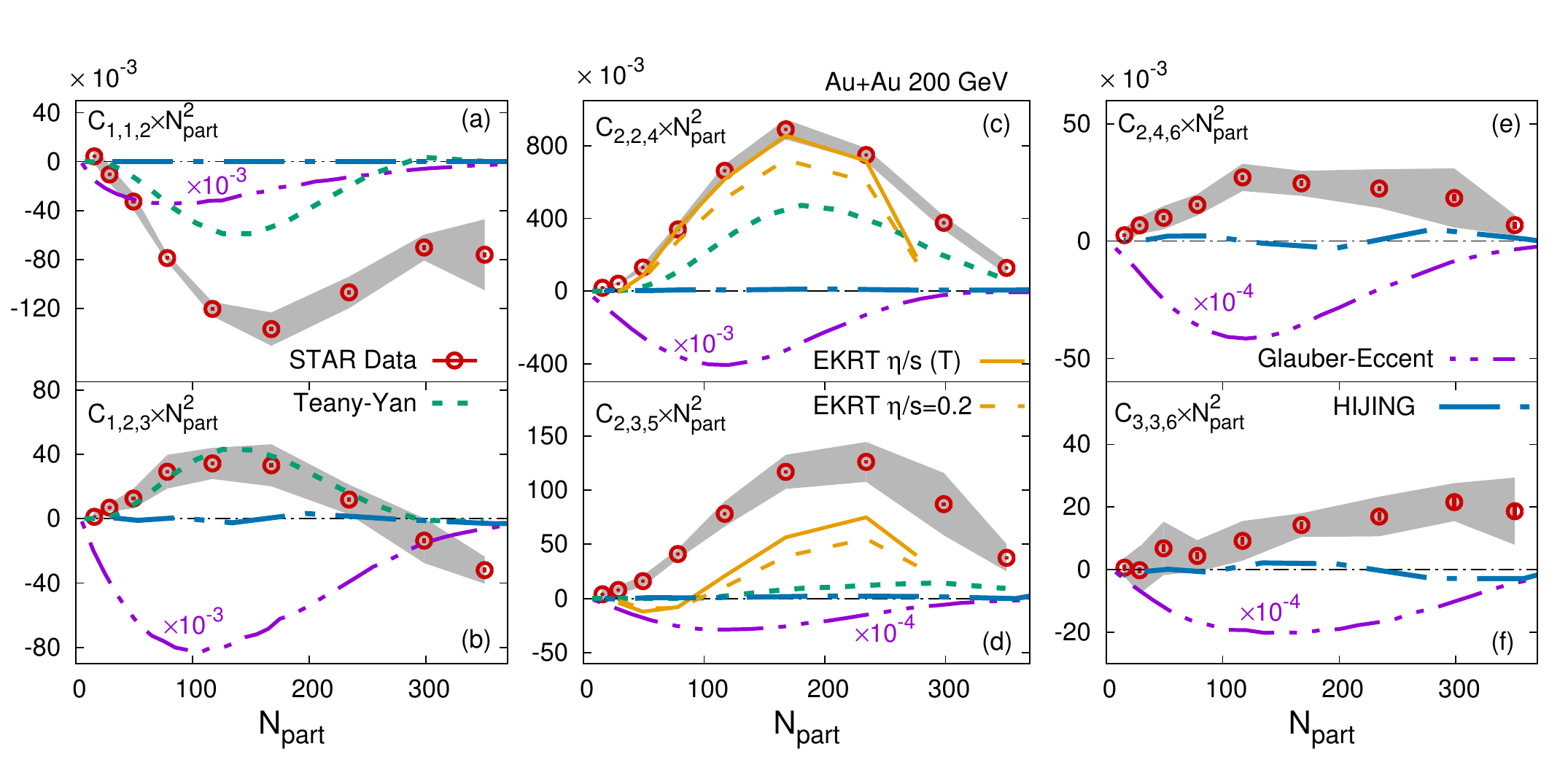}
\caption{(color online) Centrality dependence of mixed harmonic correlators $C_{m,n,m+n}$ compared to different theoretical calculations from Refs.~\cite{Teaney:2010vd,Teaney:2013dta, Niemi:2015qia, Schenke:2012wb, Schenke:2010nt}. The statistical and systematic uncertainties are shown by error bars and grey bands respectively.}
\label{fig_cosmn}
\vspace{-10pt}
\end{figure*}

In Fig.~\ref{fig_cosmn} we show centrality dependence of 
$\Delta\eta$-integrated $C_{m,n,m+n}$. 
We multiply the quantity $C_{m,n,m+n}$ by $N_{\rm part}^{2}$ 
to account for the natural dilution of correlations expected from superpositions of independent sources. 
We find that HIJING model predicts a magnitude of three-particle correlations that is consistent with zero for all harmonics. We also estimate the expectations for $C_{m,n,m+n} \approx {\la \varepsilon_m \varepsilon_n \varepsilon_{m+n} \cos(m \Phi_m + n \Phi_n - (m+n) \Phi_{m+n}) \ra}$ from purely initial state geometry using a Monte-Carlo Glauber model~\cite{Schenke:2014tga}. We find that the Glauber model 
predicts negative values for all combinations of $C_{m,n,m+n}$~\footnote{Our calculations are consistent with the estimation of plane correlations performed in Ref.~\cite{Teaney:2013dta}}. 
Since only a fraction of the initial state geometry is converted to final state anisotropy, {\it i.e.}, $v_{n} \lesssim 0.1 \times \varepsilon_{n}$ ~\cite{Teaney:2012ke}, one therefore expects ${\la v_m v_n v_{m+n} \cos(m \Psi_m + n \Psi_n - (m+n) \Psi_{m+n}) \ra} \lesssim 10^{-3} \times {\la \varepsilon_m \varepsilon_n \varepsilon_{m+n} \cos(m \Phi_m + n \Phi_n - (m+n) \Phi_{m+n}) \ra} $, we therefore scale the Glauber model calculations by factors of $\sim10^{-3}\!-\!10^{-4}$ to make a consistent data to model comparison~\cite{Teaney:2012ke}. 

\begin{figure*}[t]
\includegraphics[trim={0.5cm 0.5cm 0.5cm 0.5cm}, width=0.8\textwidth]{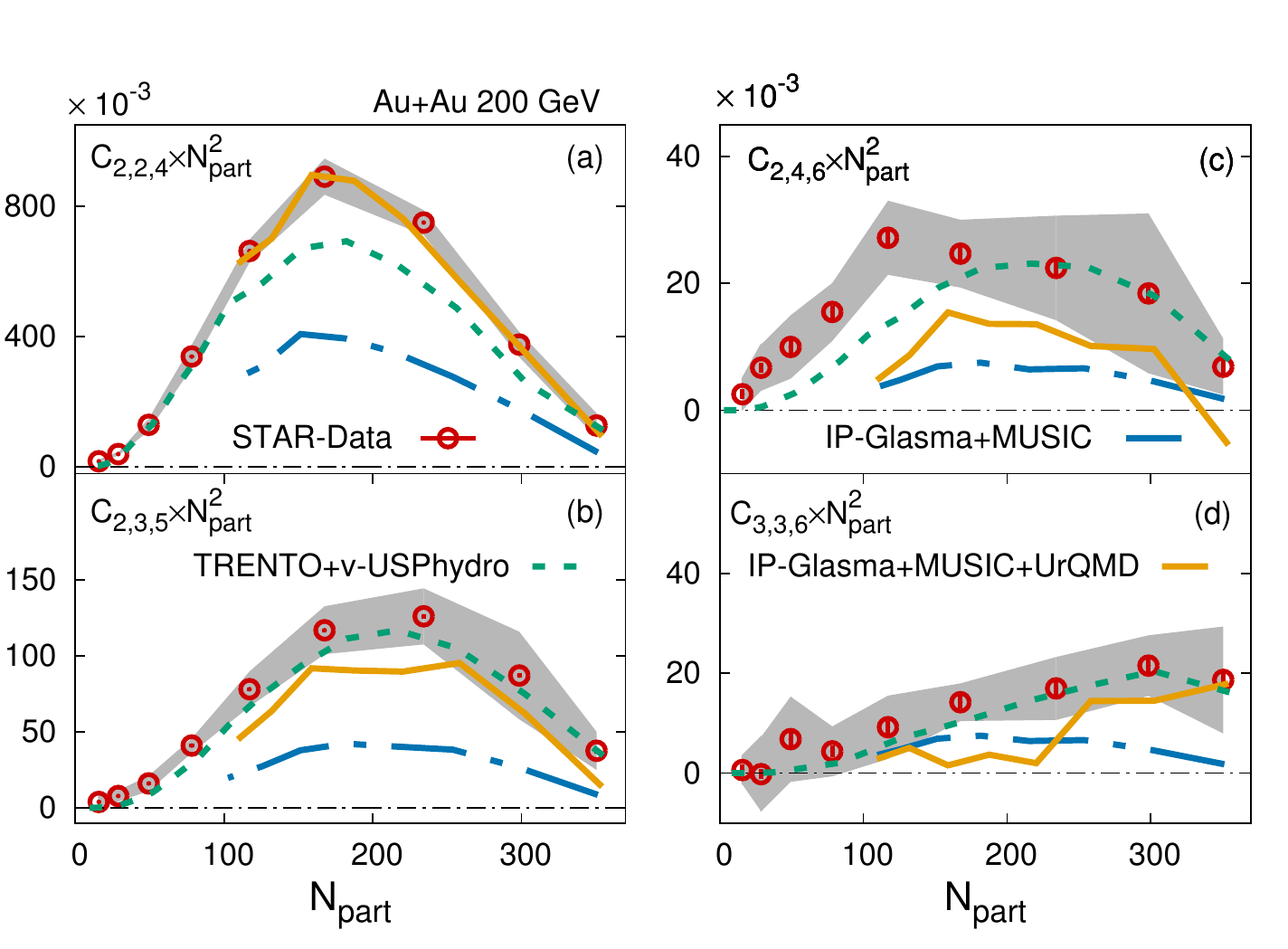}
\caption{(color online) Centrality dependence of the higher order correlators $C_{m,n,m+n}$ $(m>1)$ compared to TRENTO+v-USPhydro model calculations (shown by green dashed lines) and IP-Glasma+MUSIC calculations with and without hadronic transport using UrQMD model (shown by solid yellow and dashed blue curves).}
\label{fig_cosmn2}
\vspace{-10pt}
\end{figure*}

We compare our results with three different boost-invariant hydrodynamic model calculations that have been constrained by the global data on azimuthal correlations available so far at RHIC and the LHC. The models include : 1) 2+1 dimensional  hydrodynamic simulations with $\eta/s=1/4\pi$ with MC-Glauber initial conditions by Teaney and Yan~\cite{Teaney:2010vd,Teaney:2013dta}, 2) hydrodynamic simulations MUSIC with boost invariant IP-Glasma initial conditions~\cite{Schenke:2012wb, Schenke:2010nt} that include a constant $\eta/s=0.06$ and a temperature dependent bulk viscosity $\zeta/s~(T)$~\cite{Ryu:2015vwa} and UrQMD afterburner~\cite{Bass:1998ca}, 3) the perturbative-QCD$+$saturation$+$hydro based ``EKRT" model~\cite{Niemi:2015qia} that uses two different parameterizations of the viscosity with constant $\eta/s=0.2$ and temperature dependent $\eta/s~(T)$ with a minimum of $\left( \eta/s~(T)\right)_{\rm min}=1.5/4\pi$ at a corresponding transition temperature between a QGP and hadronic phase of $T_c=150$ MeV and 4) viscous hydrodynamic model v-USPhydro~\cite{Noronha-Hostler:2014dqa,Noronha-Hostler:2013gga} with event-by-event TRENTO initial conditions ~\cite{Moreland:2014oya} tuned to IP-Glasma~\cite{Schenke:2012wb}, that uses $\eta/s=0.05$, a freeze-out temperature of $T_{FO}=150$ MeV~\cite{Alba:2017hhe} and the most recent 2+1 flavors equation of state from the Wuppertal Budapest collaboration~\cite{Borsanyi:2013bia} combined to all known hadronic resonances from the PDG16+~\cite{Alba:2017mqu}. 

Correlators involving the first order harmonic $C_{1,1,2}$ and $C_{1,2,3}$ are shown in Fig.~\ref{fig_cosmn} (a) and (b). 
In Fig.~\ref{fig_cosmn} (a) we compare results to the hydrodynamic predictions by Teaney and Yan~\cite{Teaney:2010vd,Teaney:2013dta}. We note that since finite multiplicity effects, such as global momentum conservation, are not included in these calculations, comparisons presented for $C_{1,1,2}$ and $C_{1,2,3}$ are not intended for the purpose of constraining transport parameters. 

Any dipole anisotropy with respect to the second order harmonic plane will be exhibited in the correlator $C_{1,1,2}=\la \cos (\phi_a + \phi_b - 2\phi_c) \ra$. 
The negative value of $C_{1,1,2}$ observed in Fig.~\ref{fig_cosmn} (a) indicates that the dipole anisotropy arising at mid-rapidity is dominantly out-of-plane as predicted by the theoretical calculations in Ref.~\cite{Teaney:2010vd} and initial state geometry. It may also indicate a significant contribution from momentum conservation ~\cite{Pratt:2010gy,Bzdak:2010fd}. For the correlator $C_{1,1,2}$, it was explicitly shown that a combination of flow and momentum conservation gives rise to a negative contribution ($\sim -v_2/N$, $N$ being the multiplicity)~\cite{Pratt:2010gy,Bzdak:2010fd}. The models do not include such effects; therefore it is not surprising that they significantly under predict the data.

The centrality dependence of $C_{1,2,3}$ is shown in Fig.~\ref{fig_cosmn} (b). We see a nonzero correlation consistent with the illustrations in Fig.~\ref{fig_fluc}. The large positive values of $C_{1,2,3}$ in mid-central events are indicative~\footnote{In the mid-central events we find $C_{1,2,3}$ to be positive at low transverse momentum ($p_{T1}<1$GeV)~\cite{STAR:3pclong16}.}
 of the first harmonic anisotropy correlated with the triangularity as was first predicted in Ref.~\cite{Teaney:2010vd}. 
In the model, the hydrodynamic response of the medium changes both the sign and the centrality dependence and provides very good agreement with data for $C_{1,2,3}$ over a wide range of $N_{\rm part}$ except for the most central collisions. 
Interestingly in the most central collisions, the measurements of both $C_{1,1,2}$ and $C_{1,2,3}$ are nonzero and negative while the models predict nearly zero values for these correlators which might need further investigation~\cite{Longacre:2016xwm}.

We next report the measurement of the correlators $C_{2,2,4}$ and $C_{2,3,5}$ in Fig.~\ref{fig_cosmn} (c)-(d). The correlator $C_{2,2,4}\approx\la v_2^2 v_4 \cos(4(\Psi_2-\Psi_4))\ra$ measures the correlation between the second and the fourth order harmonics and the corresponding event planes. %
While the Glauber model results for the initial state are negative, both $C_{2,2,4}$ and $C_{2,3,5}$ exhibit strong positive values. This is consistent with the linear and nonlinear hydrodynamic response of the medium created at RHIC, in which the higher flow harmonics like $v_4$ is driven by both $\varepsilon_4$ and $\varepsilon_2$, as predicted by several theoretical calculations~\cite{Qiu:2012uy, Teaney:2013dta, Bhalerao:2013ina, Yan:2015jma, Qian:2016fpi}. 
This result is also qualitatively consistent with the measurement by the ATLAS collaboration at LHC~\cite{Aad:2014vba,Jia:2014jca}.  

The quantitive difference between the models and the measurement at RHIC is an important observation of the current study. In Fig.~\ref{fig_cosmn} (c), we observe that the hydrodynamic predictions by Teaney and Yan using constant $\eta/s$ significantly underestimate $C_{2,2,4}$. The predictions using EKRT with a temperature dependent $\eta/s$ are much closer to the data; the same using constant $\eta/s$ under predict data by about $20\%$. 
A similar trend is also observed for $C_{2,3,5}$ shown in Fig.~\ref{fig_cosmn} (d). Although all hydrodynamic models shown in this figure predict correct qualitative trends of the centrality dependence, they all significantly underestimate the magnitude of $C_{2,3,5}$. %
Such discrepancy for EKRT has been argued~\cite{Eskola:2017imo} to be related to large off-equilibrium correlations which depend on the details of the parameterization $\eta/s~(T)$. The current data will therefore provide
important constraints for the transport parameters involved in the hydrodynamic modeling at RHIC energies.

In Fig.~\ref{fig_cosmn} (e)-(f) we present the centrality dependence of 
$C_{2,4,6}$ and $C_{3,3,6}$.
Once again the positive values for $C_{2,4,6}$ and $C_{3,3,6}$, in contrast to the Glauber prediction of negative values for the initial state, indicate the importance of the nonlinear hydrodynamic response.  The EKRT predictions are not available for these correlators, it will be interesting to see if such calculations can describe the data in future. 

We revisit the centrality dependence of higher order correlators ($n>2$) in Fig.~\ref{fig_cosmn2}. Here, we compare the data with most recent hydrodynamic model calculations. The IP-Glasma + MUSIC simulations with constant $\eta/s$, tuned to global data on $v_n$s, qualitatively reproduce the trend; however they under predict the magnitude of the correlation. The IP-Glasma + MUSIC + UrQMD simulations, that include additional hadronic rescatterings, seems to be much closer to the data. This is indicative of the fact that a large fraction of the mixed-harmonic correlation is developed in the hadronic phase below a temperature of $T=165$ MeV. The addition of hadronic transport effectively increases the viscosity at lower temperature ($T < 165$ MeV)~\cite{Ryu:2015vwa}. This indicates that current data can constrain the temperature dependent transport at RHIC energies. In Fig.~\ref{fig_cosmn2} our data is also compared to the TRENTO+v-USPhydro model calculations. Although this model does not include hadronic transport, as discussed in Ref~\cite{Alba:2017hhe}, it effectively introduces a different viscous effect by choosing a lower freeze-out temperature $T_{FO}=150$ MeV, additional resonances and a different equation of state (speed of sound), as compared to IP-Glasma + MUSIC + UrQMD simulations. A reasonable description of $C_{2,3,5}$, $C_{2,4,6}$ and $C_{3,3,6}$ is obtained from the TRENTO+v-USPhydro model. In the case of $C_{2,2,4}$ the data are $20\%$ higher, which will provide further constraints for the TRENTO+v-USPhydro model~\cite{Alba:2017mqu}. It will be also interesting to see other hydro calculations by using the most recent equation of state  like TRENTO+v-USPhydro model.

After the appearance of this preprint, an extensive study using the AMPT model was shown to provide a good description of both the $\Delta\eta$ and the centrality dependence of $C_{m,n,m+n}$ in Ref.~\cite{Sun:2017ine}. Such data-model comparisons demonstrate that the longitudinal structure of the initial state, global momentum conservation and multi-phase transport can capture the underlying dynamics that drives anisotropic flow and mixed-harmonic correlations~\cite{Sun:2017ine}.

{\it Summary : } 
We presented the first measurements of the charge inclusive three-particle azimuthal correlations $C_{m,n,m+n}=  \la \la \cos(m\phi_a + n \phi_b- (m+n) \phi_c)\ra\ra$ as a function of centrality, relative pseudorapidity and harmonic numbers $m,n$ in $\sqrt{s_{NN}} = $ 200 GeV Au+Au collisions. These measurements, provide additional information about the initial geometry, the nonlinear hydrodynamic response of the medium and provide good promise to constrain temperature dependence of $\eta/s$. The centrality dependence of $C_{1,2,3}$ for the first time reveals a possible coupling between directed, elliptic, and triangular harmonic flow, which arises from fluctuations in the initial geometry. The strong $\Delta\eta$ dependence of $C_{1,2,3}$ suggests a breaking of longitudinal invariance at odds with the assumptions in many boost invariant models. 
While variations of $C_{1,2,3}$ with $\Delta\eta$ are large, $C_{2,2,4}$ and $C_{2,3,5}$ varies by only 20\% between $\Delta\eta=0$ and 2 making them most suitable for comparison to boost-invariant hydrodynamic simulations. We therefore, compared our measurements of the centrality dependence of $C_{m,n,m+n}$ with a number of boost-invariant hydrodynamic models that are constrained by global data. Such comparisons indicate that three-particle correlations can provide important constraints on fluid-dynamical modeling, in particular the temperature dependent transport at RHIC.

{\it Acknowledgement :} We thank Gabriel Denicol, Jacquelyn Noronha-Hostler, Harri Niemi, Risto Paatelainen, Bj{\"o}rn Schenke, Chun Shen, Yifeng Sun and Li Yan for providing their model predictions and helpful discussions. We thank the RHIC Operations Group and RCF at BNL, the NERSC Center at LBNL, and the Open Science Grid consortium for providing resources and support. This work was supported in part by the Office of Nuclear Physics within the U.S. DOE Office of Science, the U.S. National Science Foundation, the Ministry of Education and Science of the Russian Federation, National Natural Science Foundation of China, Chinese Academy of Science, the Ministry of Science and Technology of China and the Chinese Ministry of Education, the National Research Foundation of Korea, GA and MSMT of the Czech Republic, Department of Atomic Energy and Department of Science and Technology of the Government of India; the National Science Centre of Poland, National Research Foundation, the Ministry of Science, Education and Sports of the Republic of Croatia, RosAtom of Russia and German Bundesministerium fur Bildung, Wissenschaft, Forschung and Technologie (BMBF) and the Helmholtz Association.
 
\bibliographystyle{apsrev4-1}
\bibliography{3pcshort}

\end{document}